# The Programmable Liquid-crystal Active Coronagraphic Imager for the DAG telescope (PLACID) instrument: On-site status update ahead of first light

Jonas G. Kühn,*[a] Laurent Jolissaint,[b] Audrey Baur,[b] Liurong Lin,[a] Axel Potier,[a] Ruben Tandon,[a] Derya Öztürk Çetni,[c] Daniele Piazza,[a] Mathias Brändli,[a] Iljadin Manurung,[a] Martin Rieder[a]

[a]Division of Space and Planetary Sciences, University of Bern, Sidlerstrasse 5, 3012 Bern, Switzerland; [b]University of Applied Sciences HEIG-VD, Route de Cheseaux 1, 1401 Yverdon-les-Bains, Switzerland; [c]Atatürk University Astrophysics Research and Application Center ATASAM, Atatürk University Campus, 25240 Yakutiye/Erzurum, Turkey

## ABSTRACT

The Programmable Liquid-crystal Active Coronagraphic Imager for the DAG telescope (PLACID) instrument is a novel high-contrast direct imaging facility that was recently delivered to the Turkish 4-m DAG telescope, with first light anticipated by the end of 2024. In a nutshell, PLACID consists in a fore-optics coronagraphic intermediate stage platform, installed in-between the TROIA XAO system and the DIRAC HAWAII-1RG focal-plane array. The PLACID project, led by a consortium of Swiss Universities contracted by the Atatürk University Astrophysics Research and Application Center (ATASAM), has passed the Delivery Readiness Review (DRR) milestone in September 2023, and was delivered to ATASAM campus facilities in March 2024. The PLACID commissioning activities with the calibration light source at the summit, on the DAG telescope Nasmyth platform, are foreseen to take place this fall, with first light scheduled to take place before the end of the year. When on-sky, PLACID will be the world's first "active coronagraph" facility, fielding a customized spatial light modulator (SLM) acting as a dynamically programmable focal-plane phase mask (FPM) coronagraph from H- to Ks-band. This will provide a wealth of novel options to observers, among which software-only abilities to change or re-align the FPM pattern in function of conditions or science requirements, free of any actuator motion. Future features will include non-common path aberrations (NCPA) self-calibration, optimized coronagraphy for binary stars, as well as coherent differential imaging (CDI). We hereby present the delivered PLACID instrument, its current capabilities, and Factory Acceptance commissioning results with relevant performance metrics.

## 1. INTRODUCTION

Observing exoplanets with direct "high-contrast" imaging (HCI) carries a tremendous and synergistic potential, as HCI allows for in-situ observing of exoplanets in their host stellar environment (e.g. to study planet formation and disk interaction processes), coincidently mitigating part of the observational biases from RV and transit detections in favor of massive close-in objects. Critically, HCI also opens the door to spectroscopic characterization of the atmospheric composition of exoplanets [1]. Yet, since the initial promising discoveries of Fomalhaut b [2] and HR8799 b,c,d,e [3] about 15 years ago, the number of new confirmed detections of planetary mass objects with HCI has not exceeded a few dozen, owing for a still large technological gap that needs to be addressed by the community. Detecting exoplanets with HCI from the ground is admittedly a daunting task, not the least because of the extreme contrast ratio ($10^{-4}$ or better) required at angular separations as small as few diffraction beam widths ($\lambda/D$ units). The upcoming class of extremely large telescopes (ELTs), bearing 25- to 40-m diameter primary mirrors, will ease some of these challenges, notably by improving sensitivity and angular resolution, but several of the aforementioned issues affecting direct imaging and coronagraphy will remain, or may even worsen. Those include residual differential atmospheric refraction, resolved nearby giant stars (a side-effect of using bigger telescopes), non-common path aberrations (NCPAs), pupil registration stability, and non-ideal segmented telescope apertures.

*jonas.kuehn@unibe.ch

Part of these aspects are inherent to the larger ELTs apertures and support structures, but notably also arise from the segmented geometry of the ELTs primary mirrors, whose merit functions may even evolve over time due to dead, defective, altered (e.g. uneven reflectivity), or missing (e.g. during servicing) individual mirror segments.

The initial concept of "adaptive coronagraphy" was originally proposed by Bourget et al. in 2012 [4], using a compressed mercury (Hg) drop to generate a Lyot coronagraph with tunable diameter occulting spot. In 2016, we proposed [5] to use liquid-crystal on-silicon spatial light modulators (LCOS SLMs, see e.g. [6,7]) as active programmable focal-plane phase masks (FPM) coronagraphs, taking advantage of their resolution often exceeding 1 Mpixel and ~10 μm-sized individual pixels, enabling sufficient sampling of the telescope point-spread function (PSF) in the coronagraphic focal-plane. As a caveat, SLMs imprint a scalar phase delay pixel-wise, hence potentially limiting the contrast performance with broadband light as compared to e.g. vectorial phase plates [8,9]. Another drawback is that the use of birefringence-induced phase response requires linearly-polarized input light, thus limiting optical transmission to less than 50% for unpolarized light (but split polarization channels remain a viable option). Still, it is expected that those limitations will not represent a contrast bottleneck on-sky [10], where post-AO wavefront errors (WFE) residuals are deemed to dominate the leakage budget. On the plus side, such a SLM-based programmable FPM coronagraph will enable the observers to adapt to observing conditions (seeing, wind, etc.) in real-time, by selecting the optimal pattern in a trade-space between inner-working angle (IWA) and robustness to low-order aberrations, in particular tip/tilt jitter. This change of pattern on the SLM is entirely on the software side, with not actuator motion involved. Observers will also be able to adapt to the specificities of the science target, for example follow-up of known companions vs. blind survey, resolved vs. unresolved stars (to be relevant in the ELTs era), and in particular to multiplicity of the target. The latest scenario will make it possible to mask multiple stars in the field-of-view (FOV) of the instrument, enabling niche science case for high-contrast observations around challenging compact binary or triple stars systems of similar magnitudes [5,11]. Additional potentially interesting features of adaptive FPM coronagraphy include self-calibration of NCPAs, where the SLM can generate a Zernike wavefront sensor (WFS) pattern, similar to the VLT/SPHERE ZELDA WFS approach [12], but also its advantageous phase-shifted variant [13,14], without any other physical change in the optical path, in combination with a pupil imaging mode. Finally, it is also expected that SLMs can be used to introduce focal-plane phase diversity at specific modulation frequencies, to be able to perform time-domain coherent differential imaging (CDI) in order to dynamically disentangle coherent speckles from bona-fide incoherent off-axis astrophysical sources [14].

## 2. THE PLACID INSTRUMENT FOR THE DAG TELESCOPE

### 2.1 The PLACID concept and optical design

Back in October 2020, a consortium of Swiss universities consisting of the University of Bern and the University of Applied Sciences HEIG-VD, was awarded a procurement contract to build the high-contrast exoplanet direct imaging instrument for the new Eastern Anatolia Turkish National Observatory (DAG), a 4-m Ritchey-Chrétien telescope, located atop Karakaya Ridge (3,100-m ASL) in the Erzurum province [15]. The retained design was the Programmable Liquid-crystal Active Coronagraphic Imager for the DAG telescope (PLACID) instrument, which would serve as the first ever on-sky active and adaptive high-contrast imaging platform. The PLACID design, shown in Figure 1, consists in an optional (optically bypass-able) fore-optics intermediate coronagraphic beam train, incorporating a customized H-/Ks-band LCOS SLM from Meadowlark Corp. in a coronagraphic reflective focal-plane (off-axis angle ~ 5°). In practice, a pair of fold-mirrors (denoted PUM1 and PBM3 on Fig.1) can be lifted up to intercept the beam coming from the TROIA XAO system [16] towards the DIRAC H1RG detector [17], in order to enable the PLACID active coronagraphic imaging mode (as opposed to classical non-coronagraphic AO imaging mode). In this configuration, PLACID will slow down the telescope AO beam from F/17 to F/60, in order to ensure a PSF spatial sampling of ~10 SLM pixels per λ/D at H-band (see Table 1), and create an intermediate F/60 coronagraphic focal-plane in reflection, where the SLM panel is located. Thanks to a double-pass wire-grid linear polarizer (LP) located right in front of the SLM (not shown in Fig.1), the LCOS device will act as programmable coronagraphic FPM and diffract the unwanted stellar light, which can be blocked out downstream in the post-coronagraphic Lyot pupil-plane (LPP, see Fig.1), where a motorized filter wheel is located with a variety of Lyot masks (see Table 2). The remaining PLACID optics (all actuated) are designed to reconfigure the output beam at F/17 before propagating towards the DIRAC detector, but it is worth mentioning that an additional fold-mirror (PBM1, see Fig.1) can be actuated along the z-direction to fine-tune the output focus. A general overview of the NIR AO Nasmyth platform instrumentation configuration is presented in Figure 2.

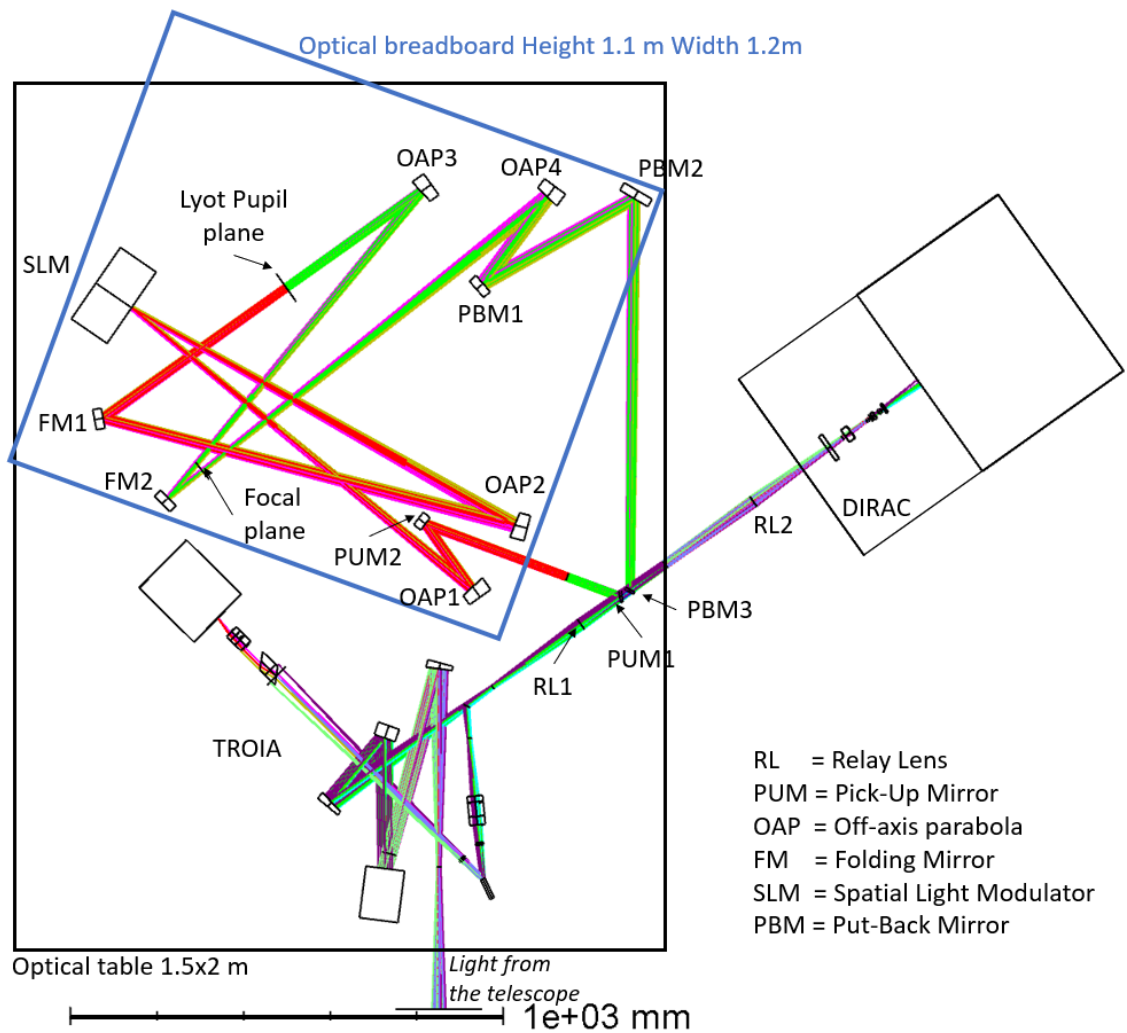

Figure 1. The PLACID Optical Zemax Design. The pair of mirrors PUM1 and PBM3 is on a motorized vertical stage, in order to enable or bypass the PLACID beam path in-between the TROIA XAO system and the DIRAC H1RG detector (choice of coronagraphic or non-coronagraphic imaging mode). The blue rectangle denotes the honeycomb optical breadboard sustaining PLACID atop the main Nasmyth optical table, the left and top corners are actually cut (see Fig.4).

**2.2 PLACID specifications and performance at Factory Acceptance**

Most key performance metrics of PLACID, as well as contractual specifications (whenever applicable), are summarized in Table 1. All measurements were obtained in HEIG-VD optical laboratory facilities, where PLACID was assembled (see Fig. 4), with a broadband source (NKT Photonics SuperK) coupled with a H-band bandpass filter (Spectrogon), and imaged with a FLI C-RED 3 InGaAs engineering camera. In this sense, the measured parameters of Table 1 are to be understood in absence of post-AO WFE and with an ideal input beam with Strehl ratio (SR) equal to unity. The relatively low throughput of PLACID – mostly due to the requirement of linearly-polarized light for the SLM – justifies in itself the existence of a "bypass" configuration between the TROIA AO and the DIRAC detector, but without PLACID, for regular AO imaging, including with angular differential imaging (ADI). Some coronagraphic imaging examples, along with post-coronagraphic raw azimuthally-averaged contrast curves are shown in Figure 3.

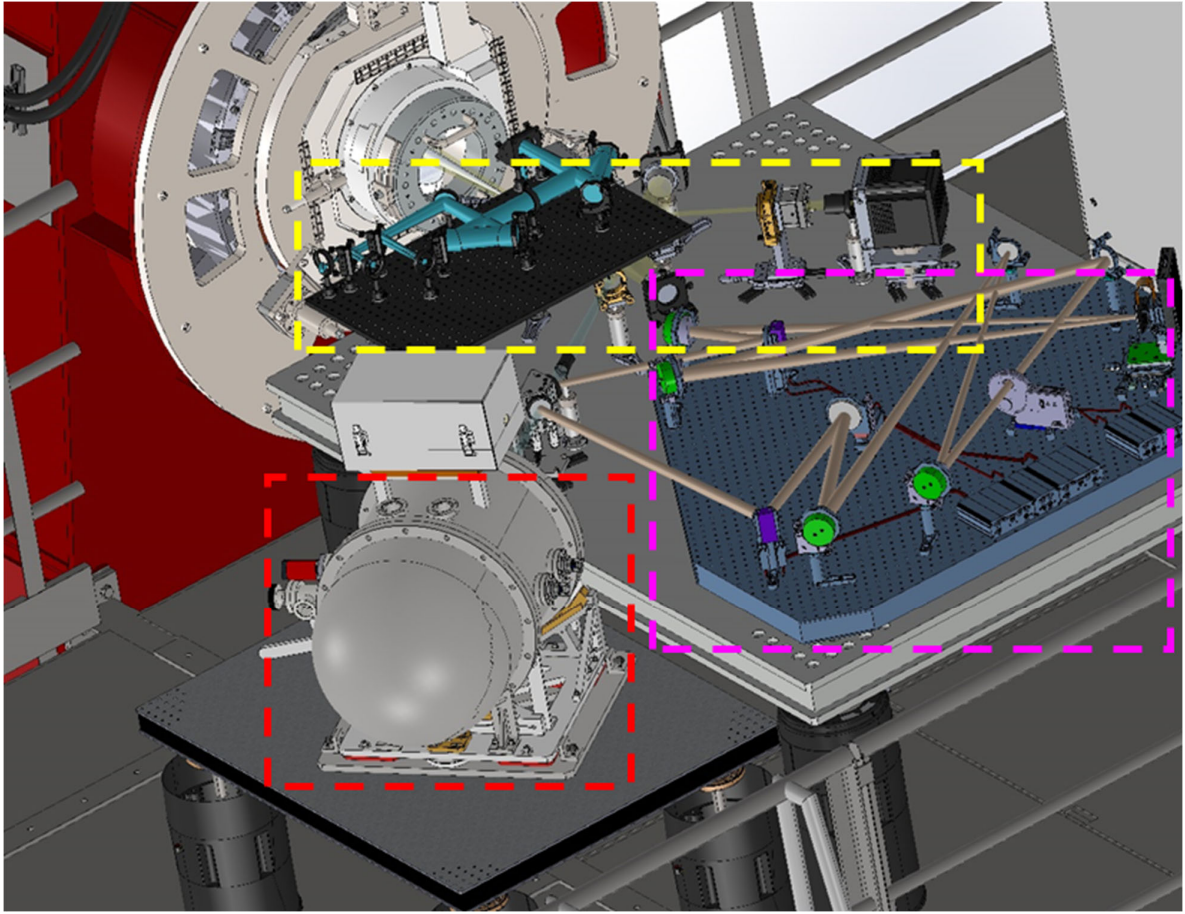

Figure 2. 3-D CAD representation of the DAG AO Nasmyth platform instrumentation configuration. The PLACID breadboard is outlined with the purple dashed rectangle, while the TROIA XAO and DIRAC detector dewar are outlined with yellow, respectively red, rectangles. The DAGOS calibration source is the thin black breadboard atop TROIA. Credits: ATASAM.

### 2.3 PLACID control software and Graphical User Interface (GUI)

The PLACID control software is Python-based and runs on a HPE DL360 dual-boot Windows/Ubuntu 1U server rack unit located in the Nasmyth computer cabinet (see Fig.8). The connection to the PLACID instrument motor controllers – and the SLM configuration channel – is done through fiber optical connection (FOC) with the use of a USB3.0 hub and a USB to FOC converter/inverter. There is an extra 10-m HDMI connection running from the PLACID PC to the SLM on the instrument, given that the LCOS panel is interfaced as an extra monitor (8-bits, 1920x1152 pixels). The graphical user interface (GUI) for PLACID was developed in-house by ATASAM, upon specifications provided by the PLACID consortium. The general layout is provided in Figure 5, and is essentially divided into three specific interfaces, from right to left on Figure 5: (i) the motor controllers interface to actuate the stepper motors (provider: Standa) of the various actuated optical mounts (incl. homing, reset to default position etc.); (ii) the motorized Lyot pupil (Thorlabs FW102C) wheel slot selection (see available pupil masks in Table 2); (iii) the SLM interface to generate, upload and save the various programmable FPM coronagraphic 8-bits greylevel patterns, but also to fine tune some other parameters like x-y digital centroiding, azimuthal rotation, phase re-scaling, topographic charge (for vortex) etc. Some functionalities will be only available in "admin mode" in order to facilitate operations for non-expert observers. It is important to stress that frame acquisition, including the fits file header structure, will be done with the DIRAC H1RG control software developed by ATASAM, hence the PLACID GUI is not the interface for frame grabbing.

Table 1. PLACID specifications and measured performance metrics at Factory Acceptance (FA) in December 2022.

| Tag | Parameter | Specification | Value (at FA) | Comments |
|---|---|---|---|---|
| BAND | Operating wavelength λ | Baseline: 1.63 um (H-band) Goal: 2.15 um (Ks-band) | H-band at first light. Ks-band planned later on best effort basis with DIRAC. | Compatible SLM with A/R coating and $2\pi$ stroke up to Ks-band. C-RED3 engineering camera can only do H-band. Thermal background at Ks is unknown. |
| SLM | Spatial Light Modulator panel | LCOS panel suitable for adaptive FPM coronagraphy | Customized 1192x1152 NIR SLM (Meadowlark) | SLM resolution: 1920 x 1152 pixels Pixel pitch: 9.2 μm Phase resolution: 8-bits (256 levels) |
| FWHM | λ/D at H-band | - | 85 mas | 115 mas at Ks-band |
| SAMP | # of SLM pixels per λ/D in focal-plane | Suitable for adaptive FPM coronagraphy | 10.6 pixels per λ/D (at H-band) | Set by F/60 f-ratio of PLACID. (13.2 pixels per λ/D at Ks-band) |
| FOV | On-sky field-of-view | > 7'' x 7'' | ~ 15.3'' x 9.2'' | SLM resolution is 1920 x 1152 1 SLM pixel ~ 8 mas on-sky |
| RES | SLM phase resolution | Suitable for adaptive FPM coronagraphy | ~ 6.4 deg / 0.11 rad (at H-band) | Set by 8-bits grey level phase resolution of SLM. |
| NULL | On-axis stellar null | 10:1 or better | > 20:1 | Achieved with H-band source at FA (VC2 FPM) |
| C2LD | Contrast at 2 λ/D (raw) | $< 5 \cdot 10^{-2}$ | $< 1.1 \cdot 10^{-2}$ | Achieved with H-band source at FA (VC2 FPM). Azimuthally-averaged. |
| C5LD | Contrast at 5 λ/D (raw) | $< 2 \cdot 10^{-3}$ | $< 5.3 \cdot 10^{-4}$ | Achieved with H-band source at FA (VC2 FPM). Azimuthally-averaged. |
| IWA | Inner-working angle | < 2.5 λ/D | < 2 λ/D | Achieved with H-band source at FA (VC2 FPM) |
| OT | Optical throughput | > 10 % | ~ 22 % | Not taking account coronagraphic transmission (FPM & Lyot stop). |

## 3. STATUS UPDATE AND NASMYTH COMMISSIONING

The PLACID project passed Factory Acceptance in Yverdon, Switzerland, in January 2023, in the laboratory settings of Figure 4. The instrument was packed for air delivery to Erzurum, Turkey, in the summer of 2023, and PLACID passed Delivery Readiness Review in September 2023 (Fig. 6-Left, Center). Following transport and customs documentation preparation through the winter of 2023/2024, the PLACID instrument was finally delivered to ATASAM facilities on Erzurum Atatürk University campus in March 2024. During the same month, PLACID shipment integrity was validated, and the instrument successfully passed Delivery Inspection Review (Fig. 6-Right).

Table 2. PLACID Lyot pupil wheel slot configuration available at first light, in terms of oversize/undersize factors in the Lyot pupui-plane ($X'$) vs. the original DAG input pupil dimensions ($X$). Future upgrades are possible.

| Slot | Name | Dp' / Dp | Ds' / Ds | ths' / ths | Throughput % |
|---|---|---|---|---|---|
| Lyot #1 | OPEN | 1.3 | - | - | 100 |
| Lyot #2 | CLOSE | 0 | - | - | 0 |
| Lyot #3 | SMALL CROSS | 0.98 | 1.2 | 5.5 | 90.4 |
| Lyot #4 | SMALL CROSS w/ ND10 | 0.98 | 1.2 | 5.5 | 9.4 [2] |
| Lyot #5 | LARGE CROSS | 0.95 | 1.3 | 11.1 | 80.1 |
| Lyot #6 | RODDIER CROSS | 0.98 | 1.05 | 5.5 | 92.2 |

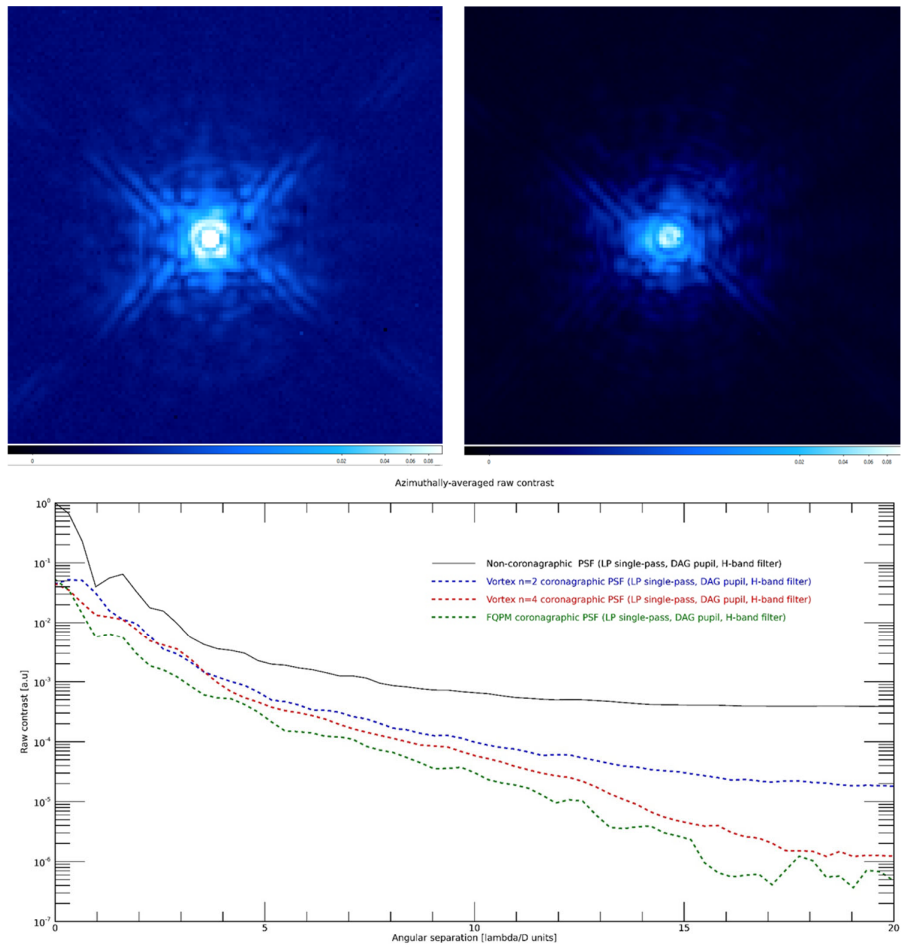

Figure 3. Examples of imaging and contrast curve results obtained with PLACID at Factory Acceptance, December 2022. (Top row, from left to right) Non-coronagraphic and coronagraphic (VC2 FPM) PSFs, normalized, same contrast scale (log); (Bottom) Examples of raw azimuthally averaged contrast curve for 3 different FPMs (VC2, VC4, FQPM).

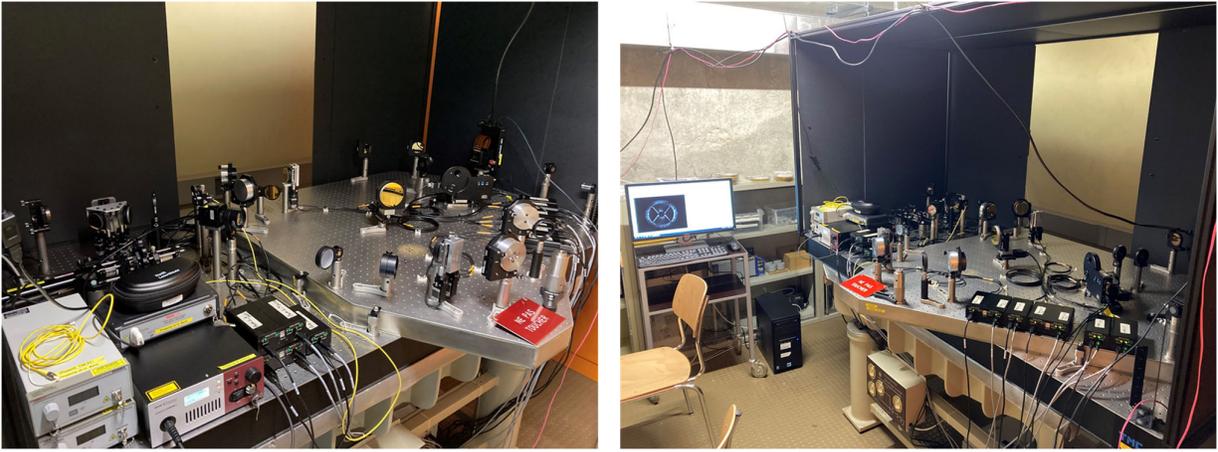

Figure 4. PLACID during Factory Acceptance Review in HEIG-VD laboratory facilities, Yverdon, Switzerland, in December 2022 / January 2023.

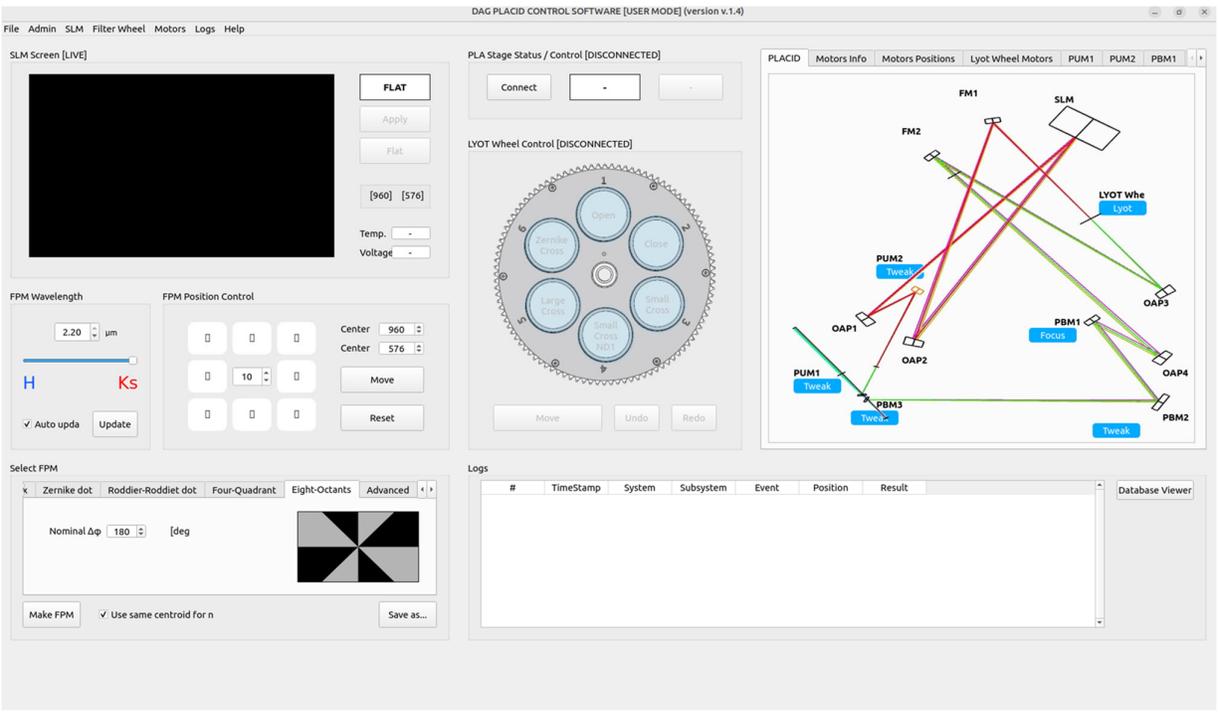

Figure 5. Preview of the Python-based PLACID GUI developed by ATASAM upon specifications of PLACID consortium.

Given that the overall DAG telescope and dome recently passed Final Acceptance as of June 2024, the entire instrumentation suit of the AO Nasmyth platform (TROIA, PLACID, DIRAC, but also the KORAY derotator [18] and the DAGOS calibration source) is scheduled for transportation to the summit through the summer of 2024. Simultaneously, also through this coming summer season, the AO Nasmyth platform will be equipped with the pressure-damped instrumentation optical tables, as well as the thermal enclosure facility (see Fig. 7). The commissioning of PLACID, as well as the other individual Nasmyth instruments, is then planned for the upcoming fall 2024 season, initially with the internal calibration sources, then DAGOS. For PLACID alone, the commissioning time is expected to last about 60 days, essentially re-assessing the performance metrics listed in Table 1. Pending successful commissioning activities of the other Nasmyth facilities, PLACID will then be fed by the TROIA beam and imaging will be performed with the DIRAC detector in place of the engineering C-RED 3 camera. Provided that those tests are compliant, and assuming that the DAG telescope infrastructure is ready, it is then realistic to expect first light around the end of 2024 or early in 2025.

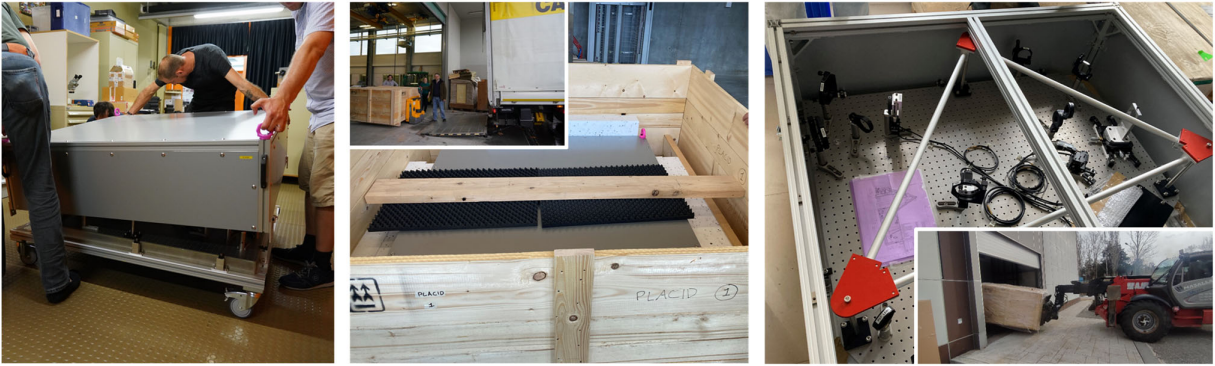

Figure 6. PLACID shipping configuration. (Left) In preparation for Delivery Readiness Review in Yverdon, Swizerland, August 2023; (Middle) Closing the dedicated PLACID crate and truck loading for air shipping to Turkey, February 2024; (Right) Unloading and Delivery Inspection Review after delivery to ATASAM facilities, Erzurum, Turkey, March 2024.

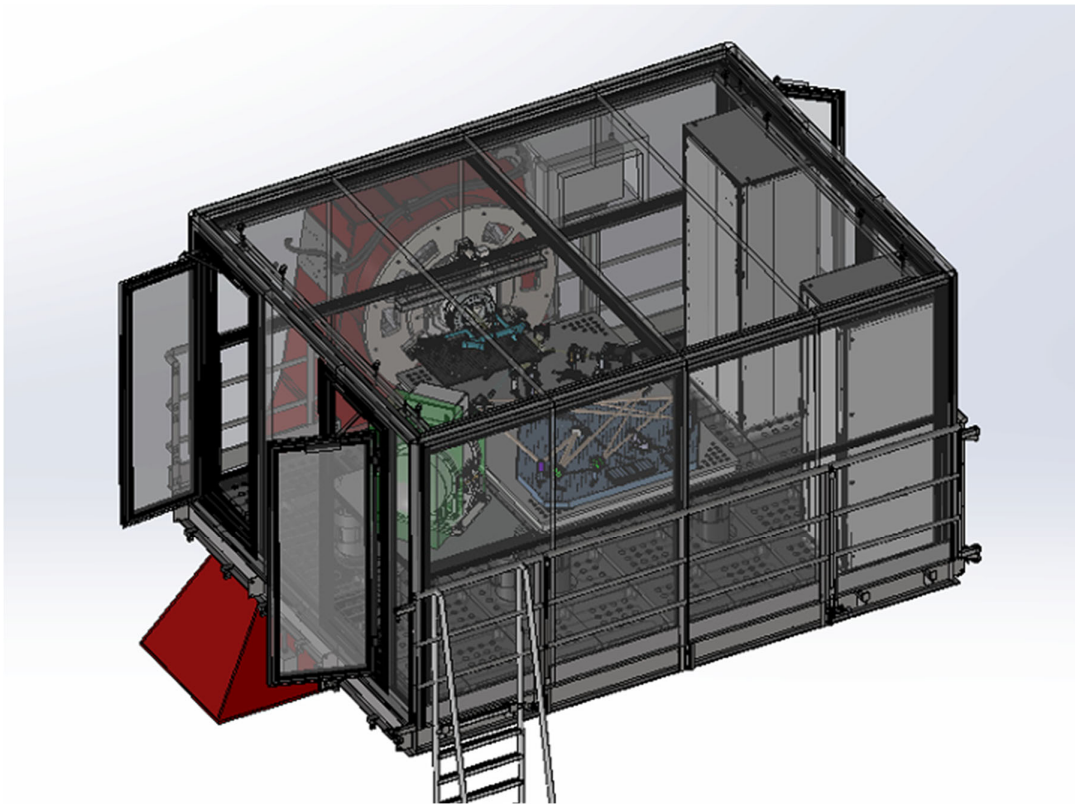

Figure 7. Current design for the DAG Nasmyth platform environment for NIR AO-assisted instrumentation, including the thermal enclosure (rendered as transparent here, but opaque in reality), scheduled for installation over the summer of 2024. Credits: ATASAM.

## 4. CONCLUSIONS AND OUTLOOK

Once on-sky later this year or by early 2025, PLACID will the first ever active programmable "all-digital" coronagraphic high-contrast imaging platform dedicated to exoplanet imaging. Beyond re-imaging of known sub-stellar companions and disks to validate the capabilities of the instrument on-sky, PLACID will be able to address a niche science discovery space around young and compact binary or triple star systems [5,11]. This latter scenario requires to develop a specific software environment for ADI active imaging of those challenging systems, to be able to synchronize the DAG Alt-Az telemetry with the display of a target-specific multi-stars coronagraphic FPM on the SLM panel at its native refresh rate of 30 Hz. Additional work planned for 2025 includes the implementation of the phase-shifting variant of the ZELDA WFS for daytime or regular calibration of NCPAs, in addition to the default phase-diversity calibration provided by the TROIA platform, as well as commissioning of the Ks-band mode of PLACID on a best-effort basis (assuming acceptable thermal background). As a versatile XAO-equipped high-contrast imaging facility on a 4-m telescope in the Northern hemisphere, PLACID should serve as an ideal platform for prototyping and collaborative on-sky validation of novel concepts of phase coronagraphy. It is also likely to become a competitive facility for follow-ups of photometric TESS and PLATO candidates, as well as GAIA DR4 and DR5 astrometric targets of interest. This will be reinforced by the fact that the amount of on-sky time for PLACID is expected to be considerable within the first few years of the DAG telescope operations, considering the relatively modest range of alternative observing modes available at the observatory. Those aspects align with one of the key missions of the PLACID project, which is to maximize the collaborative, mentoring and nurturing impact for the growing Turkish astronomical community, in particular for its young members.

Last but not least, in the 2026-2027 horizon, PLACID is foreseen to be upgraded with an optional fast imaging mode, fielding a high-speed (~400 Hz) NIR SLM and a ~1.5 kHz C-RED One eAPD camera from First Light Imaging, in the context of the ERC Rapid Active Coronagraphic Exoplanet imaging from a Ground-based Observatory (RACE-GO) project. RACE-GO aims at validating and exploiting on-sky the concepts of coherent differential imaging (CDI), at a speed exceeding the atmospheric coherent time (~10 ms in the near-infrared), in order to freeze the residual post-adaptive optics speckles. This novel observing mode will be relying on the SLM to introduce time-domain synchronous phase diversity in the focal-plane, to dynamically disentangle between coherent stellar speckles and incoherent off-axis astrophysical sources [14]. It is worth noting that the CDI imaging mode of RACE-GO will be tested on-sky well ahead of time with PLACID and the DIRAC imager, albeit at much slower speed (0.1 – 1 Hz). The availability and on-sky qualification of a novel CDI imaging mode, combined with the various other unique features of the instrument described above, is designed to make PLACID an ideal on-sky R&D prototyping platform to assist the community with the technical challenges lying ahead (e.g. ELT/PCS, NASA HWO), including with new concepts of adaptive coronagraphy optimized to non-ideal non-static segmented apertures.

## ACKNOWLEDGEMENTS

The RACE-GO project has received funding from the Swiss State Secretariat for Education, Research and Innovation (SERI), under the ERC replacement scheme following the discontinued participation of Switzerland to Horizon Europe. Part of this work has been carried out within the framework of the National Centre of Competence in Research PlanetS supported by the Swiss National Science Foundation under grants 51NF40 182901 and 51NF40 205606.